\documentclass[aps,twocolumn,pra,groupedaddress,showpacs,%
floatfix,superscriptaddress]{revtex4}

\usepackage{graphicx}
\usepackage{amssymb}
\usepackage{amsmath}
\usepackage{color}
\usepackage[normalem]{ulem}
\usepackage{subfigure} 

\begin{document}

\title{Precision spectroscopy of the 3s-3p fine structure doublet in Mg$^+$ }

\author{V.~Batteiger}
\affiliation{Max--Planck--Institut f\"ur Quantenoptik, 85748
Garching, Germany}
\author{S.~Kn\"{u}nz}
\affiliation{Max--Planck--Institut f\"ur Quantenoptik, 85748
Garching, Germany}
\author{M.~Herrmann}
\affiliation{Max--Planck--Institut f\"ur Quantenoptik, 85748
Garching, Germany}
\author{G.~Saathoff}
\affiliation{Max--Planck--Institut f\"ur Quantenoptik, 85748
Garching, Germany}
\author{H.~A.~Sch\"{u}ssler}
\affiliation{Department of Physics, Texas A\&M University,
College Station, Texas 77843, USA}
\author{B.~Bernhardt}
\affiliation{Max--Planck--Institut f\"ur Quantenoptik, 85748
Garching, Germany}
\author{T.~Wilken}
\affiliation{Max--Planck--Institut f\"ur Quantenoptik, 85748
Garching, Germany}
\author{R.~Holzwarth}
\affiliation{Max--Planck--Institut f\"ur Quantenoptik, 85748
Garching, Germany}
\author{T.~W.~H\"ansch}
\affiliation{Ludwig--Maximilians--Universit\"at M\"unchen, 80539
M\"unchen, Germany} \affiliation{Max--Planck--Institut f\"ur
Quantenoptik, 85748 Garching, Germany}
\author{Th.~Udem}
\affiliation{Max--Planck--Institut f\"ur Quantenoptik, 85748
Garching, Germany}

\date{\today}

\begin{abstract}
We apply a recently demonstrated method for precision spectroscopy on strong transitions in trapped ions to measure both fine structure components of the 3s-3p transition in $^{24}$Mg$^+$ and $^{26}$Mg$^+$. We deduce absolute frequency reference data for transition frequencies, isotope shifts and fine structure splittings that are in particular useful for comparison with quasar absorption spectra, which test possible space-time variations of the fine structure constant. The measurement accuracy improves previous literature values, when existing, by more than two orders of magnitude. 
\end{abstract}


\maketitle

\section{Introduction}
\label{section1}

\begin{figure}[b!]
\includegraphics[width=0.7\columnwidth]{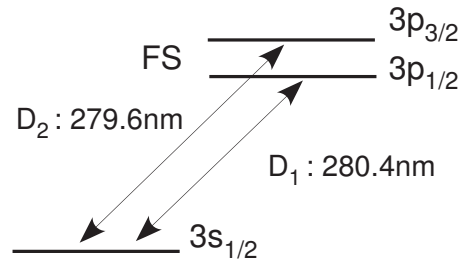}
\caption{The Mg$^+$ resonance doublet near 280 nm is isoelectronic to the 3s-3p doublet in Na historically labeled as D$_1$ and D$_2$ Fraunhofer lines. From the D$_1$ and D$_2$ transition frequencies measured in $^{24}$Mg$^+$ and $^{26}$Mg$^+$ we deduce isotope shifts and fine structure splittings (FS). The measurement accuracy allowed to resolve the isotope shift of the fine structure splitting for the first time.}
\label{fig:Mg_levels}
\end{figure}

The Mg$^+$ 3s-3p fine structure doublet near 280nm (see fig.~\ref{fig:Mg_levels}) is a prominent feature in many astronomical spectra. It can be observed by ground based telescopes, once the cosmological redshift translates the strong UV lines into the transparency range of the atmosphere. This makes the transitions good probes of physics at early epochs. One example for their astrophysical importance is the ongoing quest for a cosmological space-time variation of the fine structure constant~$\alpha$. An investigation of narrow absorption lines from gas clouds which are back-illuminated by distant quasars (quasi-stellar objects, QSO), allows to compare the value of~$\alpha$ in distant regions of the universe with the value measured on earth~\cite{Bahcall1967}. The detection of a variation of the electromagnetic coupling constant~$\alpha$ would provide compelling evidence of physics beyond the standard model.

The Mg$^+$ doublet was first used in 1976~\cite{Wolfe} to constrain variations of~$\alpha$ via the alkali doublet method~\cite{Savedoff1956}. An analysis of fine structure doublets allows to distinguish between a variation in $\alpha$ and a frequency change due to the cosmological redshift, since the ratio between the fine structure and gross structure of atomic energy levels scales as~$\alpha^2$ in leading order. Dzuba et al.~introduced the many multiplet (MM) method in 1999~\cite{Dzuba99}, a refined approach which utilizes the fact that the transition frequencies itself contain relativistic corrections $\propto\alpha^2$. The relativistic corrections are most pronounced for the ground states, so the comparison of transitions in different atomic species from the same object can significantly improve the sensitivity.
 
Webb and coworkers applied the MM method to Fe$^+$ transitions measured against the Mg$^+$ doublet and found indications that~$\alpha$ was smaller than the present day value for redshifts~$z\cong1$ (about $8\times10^9$~years ago,~$z=(\nu_{emit}-\nu_{obs})/\nu_{obs}$)~\cite{Webb99}. The same group extended their studies to larger samples and further transition lines~\cite{Murphy2001,Murphy2003}, which finally resulted in a statistically significant signature of a smaller fine structure constant $\Delta\alpha/\alpha=(-0.57\pm0.11)\times10^{-5}$. Hereby, 143 QSO absorption samples where evaluated (77 of which were low-$z$ Mg/Fe systems~\cite{Murphy2004}), all observed by the Keck observatory using the HIRES spectrometer~\cite{keckvlt}. Attempts to reproduce the results with 23 Mg-based samples observed by a different telescope and spectrometer (VLT/UVES~\cite{keckvlt}) suggest a drift consistent with zero~$\Delta\alpha/\alpha=(-0.06\pm0.06)\times10^{-5}$~\cite{Srianand2004}. This work and further investigations consistent with zero drift are summarized and disputed in~\cite{Murphy2008}. Recent ''Single Ion Differential $\alpha$ Measurement'' (SIDAM) studies on Fe$^+$ lines also gave a result compatible with zero drift~\cite{Molaro2008}. Next generation telescopes~\cite{Molaro2007}, possibly in combination with absolute frequency calibration~\cite{Murphy2007,Steinmetz2008,Braje2008,Li2008} may help to clarify the discrepancies on the reported uncertainty levels. 

Comparison with QSO absorption spectra requires laboratory reference data with an accuracy of better than 10$^{-4}$\AA~\cite{Berengut2006}, which corresponds to about 40~MHz at 280~nm. Absolute frequency calibrated measurements~\cite{comb} in the favorable environment of an ion trap, as reported in this article, provides reference data~\cite{Wolf2008,Wolf2009,Herrmann2009} that may be regarded as exact for comparison with astrophysical spectra. Besides high accuracy it is particularly important to provide isotopically resolved data. Magnesium has three stable isotopes with mass numbers 24, 25 and 26 (natural abundance 79:10:11). Different isotope abundances in the investigated astronomical objects were identified as an important source of systematic errors in the MM analysis, moreover, an investigation of the evolution of isotope abundances itself is of interest~\cite{Murphy2001,Kozlov2004,Ashenfelter2004,Fenner2005}. For these reasons an isotope shift measurement on the D$_1$ transition was requested in~\cite{Berengut2006}. Also comparison with isotope shift theory demands refined reference data since the accuracy of theoretical predictions exceeded the measurement uncertainty \cite{berengut2003,Drullinger1980}. 

So far the only isotopically resolved measurement was provided by Drullinger et al.~\cite{Drullinger1980}, where the $^{26,25,24}$Mg$^+$ isotope shifts of the D$_{2}$ line were determined with an uncertainty of 100~MHz. The most accurate literature values for the D$_{1}$ and D$_{2}$ transition frequencies originate from Fourier transform (FT) spectrometry using a hollow-cathode discharge lamp here an uncertainty of 60~MHz was reported for unresolved lines of natural isotopic abundance~\cite{Pickering1998,Pickering2000,Aldenius2006}. MM~analysis used laboratory reference data for the individual isotopic (and hyperfine structure) components~\cite{Murphy2003} that are derived from the unresolved FT measurements~\cite{Pickering1998,Pickering2000,Aldenius2006} in combination with the D$_{2}$ isotope shift results~\cite{Drullinger1980}. Our measurement of the D$_{1}$ and D$_{2}$ component of the $^{24}$Mg$^+$ and $^{26}$Mg$^+$ isotopes, improves the transition frequencies by more than two orders of magnitude. From these four absolute frequency measurements we derive $^{26-24}$Mg$^+$ isotope shifts. The isotope shift of the D$_{1}$~transition has not been measured before and the D$_{2}$ value is improved by three orders of magnitude. The accuracy of our measurement allows to resolve the $^{26-24}$Mg$^+$ isotope shift of the fine structure splitting. From the $^{24}$Mg$^+$ and $^{26}$Mg$^+$ measurements we additionally deduce $^{25}$Mg$^+$ transition frequencies, including the hyperfine structure, with an accuracy that is better than requested in~\cite{Berengut2006}.               

\begin{figure}[tb]
\includegraphics[width=1\columnwidth]{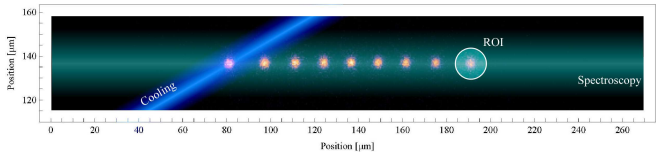}
\caption{Basic idea. A tightly focused laser beam cools ions at one side only. A weaker spectroscopy beam is applied on axis to be insensitive to possible radial micro motion. An imaging photo detector collects the essentially background free signal from single, sympathetically cooled ions.}
\label{Chain}
\end{figure}

Finally, we note that strong constraints on \emph{present-day} variations of~$\alpha$ are set terrestrially by a comparison of optical clocks based on different atomic transitions. Though optical clocks may only be compared on time scales of years, the tremendous accuracy allows to constrain variations of~$\alpha$ about an order of magnitude more accurately than MM studies, if a linear drift over 10$^{10}$~years look back time is assumed. Recent measurements of an Al$^+$ clock against a Hg$^+$ clock at NIST suggested ${\dot{\alpha}}/{\alpha}=(-1.6\pm2.3)\times10^{-17}$/year~\cite{Rosenband2008}.

\section{Method}
\label{section2}

\begin{figure}[tb]
\includegraphics[width=1\columnwidth]{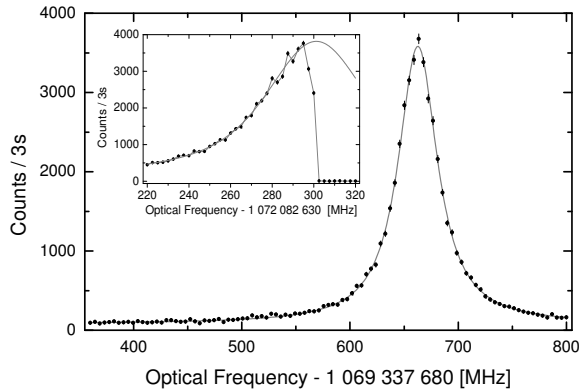}
\caption{Wide scan across the $^{24}$Mg$^+$ D$_1$ resonance. Fluorescence photons were collected for 3s per data point, the scan was randomized in order. For our measurements we restricted the tuning range to 180 MHz around the line center. Inset: For comparison we show an asymmetric $^{24}$Mg$^+$ D$_2$ line profile scanned from the low frequency side, overlayed is a Voigt fit to the rising edge, which is centered around the measured transition frequency. Here the ion was essentially uncooled, only a weak cooling beam was superimposed to prevent ion loss during the scan time.}
\label{fig:widescan2}
\end{figure}
 
 \begin{figure*}[tb]
\includegraphics[width=1.6\columnwidth]{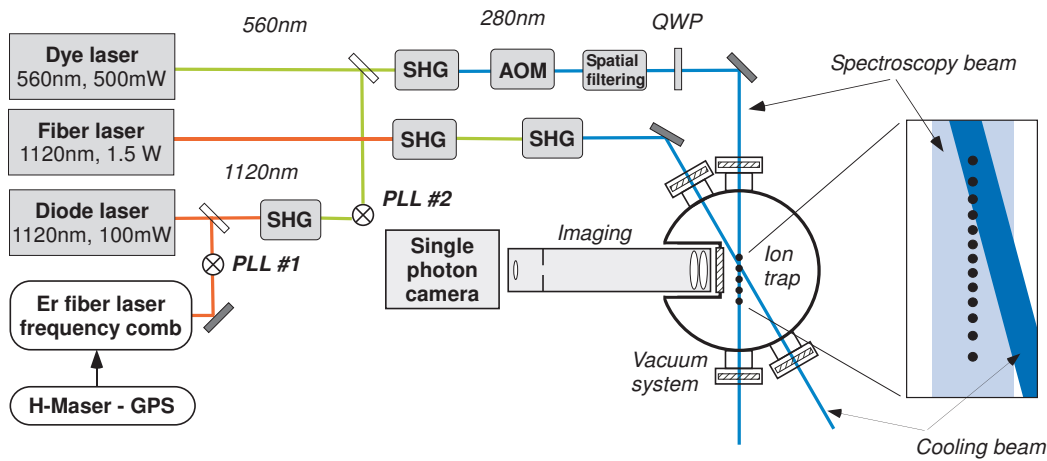}
\caption{Schematic experimental setup. All involved lasers are phase locked (PLL) to a GPS disciplined hydrogen maser. The phase coherent link between the RF reference and the UV regime via a fiber laser frequency comb and a transfer diode laser is shown for the spectroscopy dye laser. The spectroscopy beam is intensity stabilized with an AOM and spatially filtered with a 15 $\mu$m pinhole. A quarterwave plate (QWP) rotates the polarization state of the spectroscopy beam. D$_2$ cooling light is provided by a frequency quadrupled fiber laser~\cite{Friedenauer2006}. An alternative setup generating both beams from a single laser output is shown in~\cite{Herrmann2009}.}
\label{fig:Setup}
\end{figure*}
 
Lines observed in QSO absorption spectra are usually dipole-allowed transitions from the ground state~\cite{Berengut2006}. Spectroscopy on such transitions in trapped ions requires different spectroscopic approaches than probing narrow clock transitions. Spectroscopy in the favorable sideband resolved regime~\cite{Bergquist1987}, that allows to avoid first order Doppler and recoil shifts, is hard to establish, since typical transition bandwidths $\Gamma_{trans}$ are broad compared to practical secular trap frequencies $\omega_{sec}$. In the weak binding limit ($\omega_{sec}\ll\Gamma_{trans}$) the absorption spectrum is strongly asymmetric if no further measures are taken and the transition is probed with a single laser beam. In this regime the motional sidebands overlap and for constant temperature a Voigt profile would be observed, just as for an unbound atom~\cite{Wineland1979}. However, scattering photons changes the temperature of the ion; on the low-frequency (``red'') side of the resonance the ion is cooled to a detuning dependent temperature, on the high-frequency (``blue'') side it is heated and expelled from the trap center. Therefore the Voigt spectrum is heavily distorted: On the red side due to a detuning dependent equilibrium temperature and thus varying Gaussian contribution to the linewidth. Near resonance the temperature diverges and the ion is lost, so the blue side is not observed~\cite{Nagourney1983} (see also fig.~\ref{fig:widescan2} inset). Therefore, in view of precision spectroscopy, the crux is to suppress adverse temperature variations induced by the spectroscopy laser. Since the statistical uncertainty determining the line center is given by the linewidth divided by the signal-to-noise ratio of the data, a suitable approach should ensure low temperature (i.e.~narrow linewidth) and a high scattering rate. Previously used approaches include a two laser method~\cite{Drullinger1980}, a double resonance scheme~\cite{Wineland1981}, buffer gas cooling~\cite{Nakamura2006} and chopped detection~\cite{Wolf2008}. Our novel spectroscopy technique allows to record essentially unperturbed Voigt profiles for the first time. The method is discussed in detail in~\cite{Herrmann2009}, here we will only sketch the basic idea. We prepare a crystallized chain of ions which is continuously Doppler cooled at one side only (see fig.~\ref{Chain}). A weaker spectroscopy laser probes sympathetically cooled ions at the other end of the chain, where we collect a basically background free fluorescence signal from single ions using an imaging photon detection
system. This avoids both the background and ac Stark shift from the cooling laser. Counterintuitive to a thermal conductivity picture, the symmetric eigenmode structure efficiently couples cooling and spectroscopy ions at opposite ends of the chain, i.e.~the sympathetically cooled ions are cooled just as efficiently as directly cooled ions. A detailed analysis of the considered transitions shows: Residual temperature variations lead to systematic shifts of the line center always below $10^{-4}$ of the linewidth, as long as the cooling laser damping exceeds the heating induced by the spectroscopy beam for all detunings. Therefore, the method enables to observe virtually unperturbed resonances from single, cold ions with high signal-to-noise ratio (see fig.~\ref{fig:widescan2}).

\section{Setup, Measurement and Results}
\label{section3}

\begin{figure*}[htb]
  \centering
   \subfigure[~$^{26}$Mg$^+$~D$_2$]{\includegraphics[width=0.65\columnwidth]{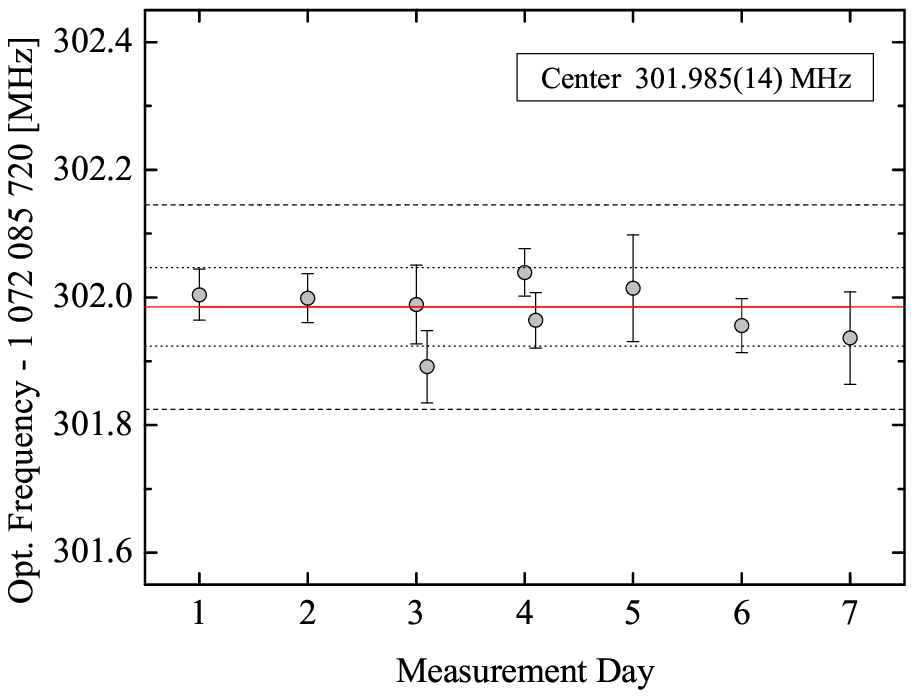}}\qquad
   \subfigure[~$^{24}$Mg$^+$~D$_1$]{\includegraphics[width=0.62\columnwidth]{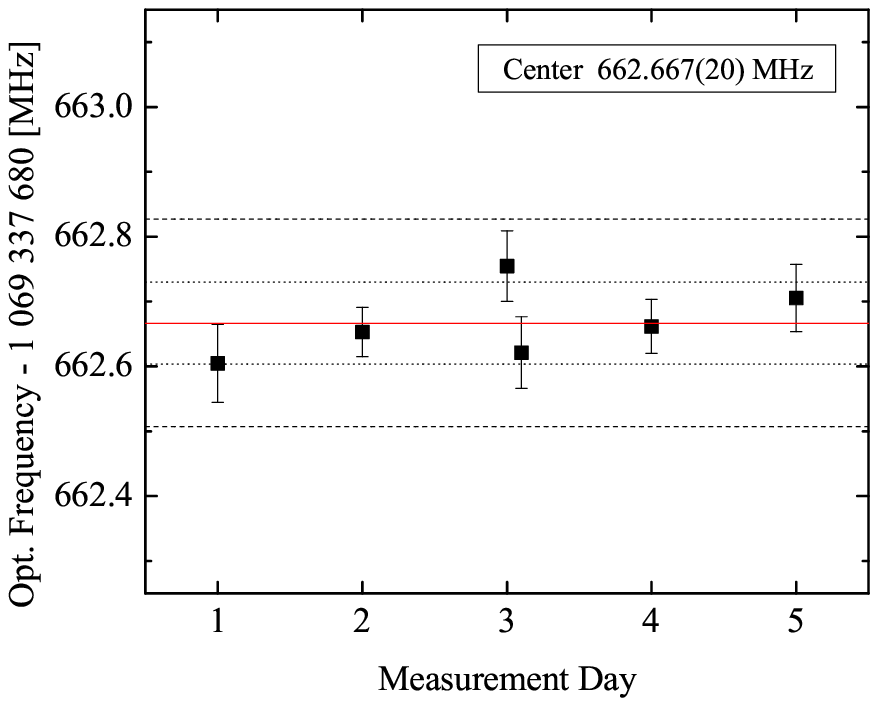}}\qquad
   \subfigure[~$^{26}$Mg$^+$~D$_1$]{\includegraphics[width=0.62\columnwidth]{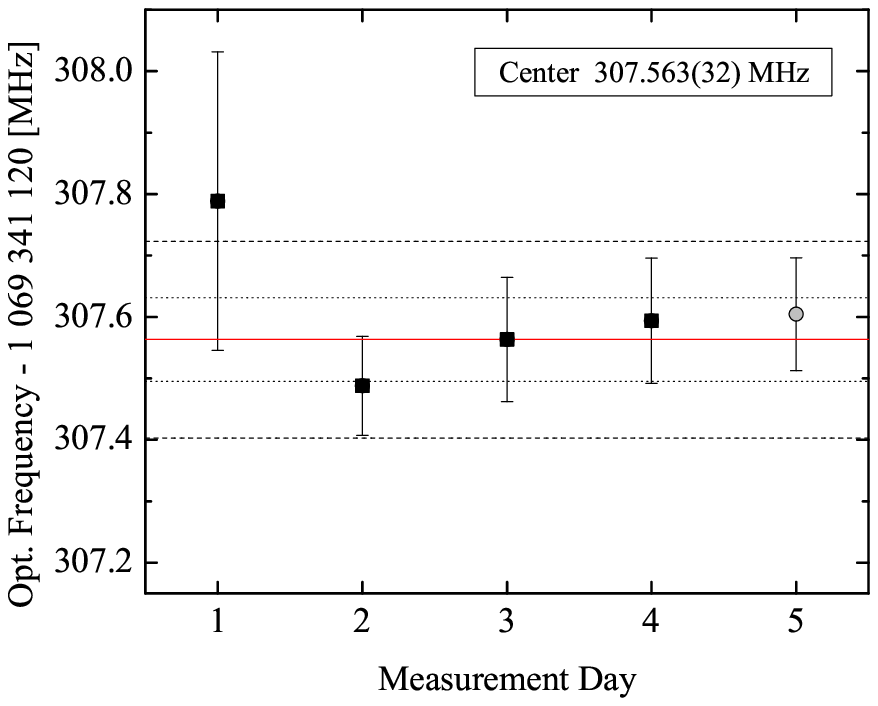}}
  \caption{Measurement of two isotopic components of the D$_1$ and D$_2$ transition. Outer dashed lines represent the error bar on the absolute transition frequencies due to correlated amplitude and phase modulation, inner dotted lines the reduced error bar for the relative determination of isotope shifts and fine structure splittings. Squares indicate measurements with two laser systems according to fig.\ref{fig:Setup}, circles measurements with the double pass AOM scheme described together with the measurement of the $^{24}$Mg$^+$~D$_2$ component in ~\cite{Herrmann2009}.}
  \label{fig:Data}
\end{figure*}

The experimental setup is described in~\cite{Herrmann2009}, here we summarize basic features and illuminate further aspects in greater detail. The Mg$^+$ ions were stored in a linear Paul trap operated at 15.8~MHz. The radial secular frequencies were about 1~MHz, while the axial secular frequency was chosen to be about 60~kHz. Ions were loaded into the trap by two-step photo ionization via the 3s$_0$-3p$_1$ transition in neutral Mg with a frequency doubled dye laser at 285~nm~(Coherent 699). References~\cite{Kjaergaard2000,Gulde2001,Lucas2004,Steele2007,Tanaka2007} showed that isotope selective photo ionization is possible when atom and laser beams are carefully arranged. Our trap was loaded from neutral Mg vapor of natural abundance (Doppler broadening: about 3~GHz, $^{26-24}$Mg~3s$_0$-3p$_1$ isotope shift: 1.4~GHz~\cite{Salumbides2006}). Yet we observed purified chains of up to 30 $^{26}$Mg$^+$ ions, when we applied additional radio frequencies resonant to the $^{24}$Mg$^+$ and $^{25}$Mg$^+$ radial secular frequencies to an electrode parallel to the trap axis during the loading period. Cooling and spectroscopy light was either provided by another frequency doubled dye laser (Coherent 699) or a frequency quadrupled Yb fiber laser system~\cite{Friedenauer2006} at 1118~nm (Koheras Boostik), which could only interact with the D$_2$ transition (90~GHz tuning range). A schematic setup using both lasers is shown in fig.~\ref{fig:Setup}. After a failure of the fiber laser the cooling and spectroscopy beams were produced by the frequency doubled dye laser. In this case the UV output was split and passed through two double pass acousto-optic modulators  (AOM) setups~\cite{DPAOM}, which allowed to independently control the frequency and amplitude of both beams. The intensities were set to $s_{cool}\sim0.5$ and $s_{probe}\sim7\times10^{-4}$ in units of the saturation intensity ($I_{sat}\sim$~2.47~mW/mm$^2$). The cooling beam was detuned 64~MHz below resonance, the practical tuning range of the spectroscopy laser was 180~MHz, limited by the efficiency of the scanning AOM in double pass configuration. The cooling beam was tightly focused to a beam radius of $w_0\sim20\,\mu$m and enclosed an angle of about 15$^\circ$ with the trap axis predefined by the optical access of our vacuum vessel. Given that geometry we found optimal experimental conditions when we cooled 2-3 outer ions of a chain containing 8-10 ions. This allowed us to perform low background spectroscopy on 2-3 ions at the other end of the chain. Working with shorter chains decreased the number of low background spectroscopy ions, longer chains increased the probability that a MgH$^+$ ion was generated by a chemical reaction with the background gas. Dark MgH$^+$ ions embedded in the chain changed their site on time scales of few seconds. Once they enter the cooling or spectroscopy region it turns dark which distorts the line profile. We therefore reloaded our trap once an impurity ion was observed. The spectroscopy beam collinear with the ion chain was stabilized in intensity, spatially filtered with a small pinhole and loosely focused to about $w_0\sim200\,\mu$m in the trapping region. Additional spatial filtering was necessary since tuning of the double pass AOM led to frequency dependent distortions of the beam profile, observable with a beam profiling camera. We stabilized the \emph{total} intensity, but the ions sampled frequency dependent \emph{local} intensity variations, which resulted in an apparent alignment sensitivity of the line center. Filtering with a~15~$\mu$m size pinhole removed the beam profile distortions and the alignment sensitivity of the line center. 

The RF reference for all involved frequency combs, phase locked loops and the spectroscopy AOM was provided by a hydrogen maser. A comparison between the free-running maser and the global positioning system showed a deviation smaller than $1.7\times10^{-13}$ over the whole measurement period, which is sufficient for our purpose. The phase coherent link between the RF reference and the optical regime was achieved with a 100~MHz repetition rate erbium doped fiber laser frequency comb. The frequency comb reaches well into the fourth sub-harmonic of the Mg$^+$ doublet at about 1120~nm. To bridge the gap to the green dye laser output we phase locked a diode laser at 1120~nm to a mode of the frequency comb and recorded a beat note between its frequency doubled output and the dye laser (see fig.~\ref{fig:Setup}). A synthesizer mixed the beat frequency down to a constant offset lock frequency, which allowed us to arbitrarily set the frequency of the phase locked dye laser with respect to the IR comb modes. This technique was also used to tune the spectroscopy laser, when we cooled the ion chain with the independent fiber laser system. The mode number of the IR comb, used for the measurement of the $^{24}$Mg$^+$~D$_2$ line was determined with a second frequency comb with repetition rate of 250 MHz (Menlo Systems)~\cite{Herrmann2009}. 
We deduced the mode number for the remaining transitions relative to the known $^{24}$Mg$^+$~D$_2$ component using two independent wavemeters with an relative accuracy considerably better than the repetition rate (Burleigh WA 1500, High Finesse WS7).

Every single measurement consisted of 30 - 40 frequency settings within a 180~MHz ($>$~4~$\Gamma$) wide range around the line center, fluorescence photon counts were recorded for 3~s each. The chronological order of probe laser frequencies was randomized for each line scan to be insensitive to a great variety of possible drifts. Photons were collected with a f/2 imaging system attached to a single photon camera (Quantar Mepsicron II). The spatially resolved count rate was digitized in 512~$\times$~512 pixel images. To read out the fluorescence of a single ion we set a circular region of interest around the ion using software (see fig.~\ref{Chain}). 

\begin{table*}
\caption{Absolute transition frequencies for individual isotopic components. Pickering et al.~do not explicitly specify an uncertainty for the $^{24}$Mg$^+$ component (see table~1 in \cite{Pickering1998}), for comparison we adopt the composite line uncertainty of 60~MHz.}
\begin{tabular}{|l|c|c|c|}
\hline
Measured transitions~~~~&  This work &~~~Drullinger et al.~\cite{Drullinger1980}~~~&~~~Pickering et al.~\cite{Pickering1998}~~~\\
\hline
\hline
~~$^{24}$Mg$^+$~D$_1$ (3s$_{1/2}$-3p$_{1/2}$) &~~~~1 069 338 342.56 (16) MHz~~~~& - & 1 069 338 293 (60) MHz \\
\hline
~~$^{26}$Mg$^+$~D$_1$ (3s$_{1/2}$-3p$_{1/2}$) &~~~~1 069 341 427.47 (16) MHz~~~~& - & - \\
\hline
~~$^{24}$Mg$^+$~D$_2$ (3s$_{1/2}$-3p$_{3/2}$)~\cite{Herrmann2009}~~~&~~~~1 072 082 934.33 (16) MHz~~~~&~~~1 072 082 833 (120) MHz~~~& 1 072 082 862 (60) MHz   \\
\hline
~~$^{26}$Mg$^+$~D$_2$ (3s$_{1/2}$-3p$_{3/2}$) &~~~~1 072 086 021.89 (16) MHz~~~~&~~~1 072 085 883 (156) MHz~~~& -  \\
\hline
\end{tabular}
\label{TableResults}
\end{table*}

The D$_2$ transition shows two magnetic sublevels for the s$_{1/2}$ ground state and four magnetic sublevels for the p$_{3/2}$ excited state. Magnetic fields in direction of the spectroscopy laser therefore shift the line center if the polarization is not perfectly linear. To compensate such shifts we introduced a quarter wave plate (QWP) into the beam path and measured the line center as function of the polarization, by rotating the QWP in 45$^\circ$ steps. We observed the expected sinusoidal modulation of the line center with an amplitude smaller than 950~kHz, a sinusoidal fit to the data compensates linear Zeeman shifts. Each measurement day in fig.~\ref{fig:Data}a consisted of at least one QWP rotation scan. Binning the data into measurement days serves as an important consistency check against possible systematics induced by operators and the lab environment. Since the s$_{1/2}$ and p$_{1/2}$ levels show two magnetic sublevels each, only line broadening is expected for the D$_1$ line. Nevertheless we continued with the QWP evaluation scheme to rule out unexpected polarization effects on the line center. No significant polarization dependence of the D$_1$ line was observed. 

We assume that the limiting systematic uncertainty is introduced by possible asymmetries in the rather broad spectrum of the frequency doubled dye laser, which was about 10~MHz. The phase locked loop to the frequency doubled diode laser stabilizes the carrier phases~(see fig.~\ref{fig:Setup}). However, correlated amplitude- and phase modulation (AM/PM) can lead to asymmetries in the frequency spectrum with respect to the carrier frequency. Therefore the center-of-gravity of the spectroscopy laser spectrum can shift against its absolute frequency. Since the fluorescence profile is a convolution of the atomic response and the laser spectrum, the counted frequency may differ from the true transition frequency. To quantify the impact of this effect, we measured the in-loop heterodyne beat note between the diode laser and the dye laser with a spectrum analyzer referenced to the same RF reference as the involved offset lock frequency. We fitted a Gaussian profile to the beat note and found an average deviation from the offset lock frequency of 80(60)~kHz at 560~nm, therefore we assume an uncertainty of 160 kHz in the UV, an order of magnitude larger than e.g.~the 14~kHz statistical uncertainty of the $^{26}$Mg$^+$~D$_2$ measurement. A consistency check with the frequency quadrupled fiber laser directly locked to the IR frequency comb gave good agreement within the statistical error bars~\cite{Herrmann2009}. All other considered systematic effects, including method inherent shifts due to residual detuning dependent temperature variations, are much smaller: ac Stark shift due to residual background from the cooling laser (30 kHz), dc Stark shift from the trapping fields (0.1~Hz), line shape model (270~Hz), RF reference (180~Hz) and 2nd order Doppler shift (-0.3~Hz). We corrected the recoil shift of -106~kHz for $^{24}$Mg$^+$ and -98~kHz for $^{26}$Mg$^+$ transition frequencies.

We measured the $^{26}$Mg$^+$~D$_2$ line on six measurement days (152 line scans, statistical uncertainty~14~kHz) and the $^{24}$Mg$^+$~D$_1$ line on five measurement days (119 line scans, statistical uncertainty~20~kHz). Due to technical problems we were only able to measure a total of 67 lines in five
measurement days of the $^{26}$Mg$^+$~D$_1$ component. Still, the statistical uncertainty of the measurement reached 32~kHz and all binned measurement days agree within their error bars~(see fig.~\ref{fig:Data}c). The numerical values of the transition frequencies are given in table~\ref{TableResults}.

The reproducibility of our measurement over time periods of days and months, also indicates that the systematic uncertainty due to correlated AM/PM enters rather as a constant offset on the absolute frequency, than as a time-varying scatter. We therefore assume that the uncertainty of measured isotope shift frequencies and fine structure splittings is smaller than 160 kHz due to cancellation of correlated AM/PM effects. Possible shifts due to residual detuning dependent distortions of the intensity profile induced by the scanning double pass AOM should not influence measurements where the phase locked spectroscopy laser was tuned by changing the offset lock frequency with a synthesizer. Thus, the reproducibility over time and against changes in the setup indicates that the statistical uncertainty of a typical measurement day is an appropriate upper bound on residual systematic deviations due to drifting correlated AM/PM and residual distortions of the intensity profile~(see fig.~\ref{fig:Data}). We therefore quadratically add a systematic deviation of 60 kHz to the statistical uncertainty of the involved transition frequencies, when we determine the relative quantities summarized in table~\ref{TableResultsII}.      

\begin{table} [tb]
\caption{Isotope shifts~($\delta\nu^{A',A}$), fine structure splittings (FS) and isotope shift of the fine structure splitting deduced from the D$_1$ and D$_2$ absolute frequency measurements. The error bars are reduced due to common mode rejection of systematic effects.}
\begin{tabular}{|l|c|}
\hline
~$\delta\nu^{26,24}$~D$_1$~(3s$_{1/2}$-3p$_{1/2}$)~~~& 3084.905 (93) MHz  \\
\hline
~$\delta\nu^{26,24}$~D$_2$~(3s$_{1/2}$-3p$_{3/2}$)~& 3087.560 (87) MHz \\
\hline
~$^{24}$Mg$^+$~FS~(3p$_{1/2}$-3p$_{3/2}$)~&~~~2 744 591.767 (88) MHz~~~\\
\hline
~$^{26}$Mg$^+$~FS~(3p$_{1/2}$-3p$_{3/2}$)~&~~~2 744 594.422 (92) MHz~~~\\
\hline
~$\delta\nu^{26,24}$~FS~(3p$_{1/2}$-3p$_{3/2}$)~& 2.66 (13) MHz \\
\hline
\end{tabular}
\label{TableResultsII}
\end{table}

\section{$^{25}$Mg$^+$ isotope shift and hyperfine structure}
\label{section4}

$^{25}$Mg$^+$ has a nuclear spin {\it I}~=~5/2. Its resulting hyperfine structure makes additional beams necessary to establish a cycling transition and repump from dark hyperfine levels, which were not available at the time of the measurement. Yet, having measured the $^{26-24}$Mg$^+$ isotope shifts, we may deduce $^{25}$Mg$^+$ transition frequencies with an uncertainty of about 20~MHz. The frequency shift for a transition between isotopes with atomic masses {\it A'} and~{\it A} respectively can be written as~\cite{berengut2003,Drake}  
\begin{equation}
\label{shiftformula}
\delta\nu^{A',A}=(k_{\rm{NMS}}+k_{\rm{SMS}})(\frac{1}{A'}-\frac{1}{A})+F_{FS}~\delta\left\langle r^2\right\rangle^{A',A},
\end{equation}
where $\delta\nu^{A',A}~:=\nu^{A'}-\nu^{A}$, $F_{FS}$ is the field shift constant and $\delta\left\langle r^2\right\rangle^{A',A}$ is the difference between the root-mean-square (rms) nuclear charge radii of the involved isotopes. The mass shift arising from the nuclear recoil is usually treated as the sum of the normal mass shift NMS, that can easily be determined and the specific mass shift SMS, which is difficult to evaluate accurately and has been subject to a variety of theoretical investigations \cite{berengut2003,safronova2001,tupitsyn2003,Korol2007}. The field shift arising from different nuclear charge distributions for different isotopes, turns out to be a small contribution (about 20~MHz) compared to the total mass shift of the $^{26-24}$Mg$^{+}$ D$_1$ and D$_2$ line (about 3~GHz). We express the sum of the mass shift constants in eqn.~(\ref{shiftformula}) in terms of the measured $^{26-24}$Mg$^+$ isotope shift and arrive at an expression 
\begin{equation}
\label{shiftformulaII}
\delta\nu^{25,24}=a~\delta\nu^{26,24}+F_{FS}~(\delta\left\langle r^2\right\rangle^{25,24}-a~\delta\left\langle r^2\right\rangle^{26,24}),
\end{equation}
where $a=\frac{m_{26}(m_{25}-m_{24})}{m_{25}(m_{26}-m_{24})}$, that allows to deduce the $^{25-24}$Mg$^+$ isotope shifts using field shift constant determinations $F_{FS}$  from~\cite{berengut2003,safronova2001,tupitsyn2003} and tabulated rms nuclear charge radii, which were measured by electron scattering and muonic atoms x-rays~\cite{angeli2004}. The uncertainty introduced by the measurement uncertainty of the rms nuclear charge radii dominates over the influence of different theoretical field shift constants derived in~\cite{berengut2003,safronova2001,tupitsyn2003,Korol2007}. The results are: $\delta\nu^{25,24}$Mg$^+$~D$_1$~=~1620(19)~MHz and $\delta\nu^{25,24}$Mg$^+$~D$_2$~=~1621(19)~MHz in agreement with~\cite{Drullinger1980}. 

Knowing the $^{25-24}$Mg$^+$ isotope shift (i.e.~the hyperfine centroid) one can evaluate transition frequencies between $^{25}$Mg$^+$ hyperfine structure levels. The ground state hyperfine constant was accurately measured in~\cite{Itano1981}. Magnetic dipole and electric quadrupole hyperfine constants for 3p states in $^{25}$Mg$^+$ are evaluated in~\cite{Sur2005,Safronova1998}, uncertainty estimates are not given, but the computation reproduces the experimental value of the ground state hyperfine constant within 0.6\%. The largest shift to a 3p state amounts to 180 MHz (3p$_{1/2}$,{\it F}=2), the uncertainties of the 3p hyperfine constants (assuming that they are smaller than 10\%) may therefore be neglected against the 19~MHz uncertainty in the determination of the $^{25}$Mg$^+$ hyperfine centroid, which is smaller than the transition bandwidth $\Gamma_{trans}$~=~41~MHz and smaller than the 60~MHz requested for the modeling of isotopically unresolved Mg$^+$ lines in astronomical spectra~\cite{Berengut2006}. Using hyperfine constants tabulated in~\cite{Sur2005} one can deduce the whole hyperfine structure of 3p states in $^{25}$Mg$^+$. The absolute frequency of the 3s$_{1/2}$({\it F}=3)-3p$_{3/2}$({\it F}=4) cycling transition, which is widely used in trapped ion experiments, reads for instance: 1~072~085~225~(19)~MHz. This value may be updated when improved hyperfine constants or rms nuclear charge radii become available.

\section{Comparison with literature}
\label{section5}
\subsection{Comparison with Previous Experimental Data \label{section2a}}

\begin{table} [tb]
\caption{Comparison of transition frequencies for a composition of natural isotopic abundance. All frequencies are given in MHz.}
\begin{tabular}{|l|l|l|}
\hline
Reference &~~~~~D$_1$~[MHz] &~~~~~D$_2$~[MHz]~\\
\hline
\hline
This work&~1 069 338 844.2 (1.9)~&~1 072 083 436.4 (1.9)~\\
\hline
Drullinger~\cite{Drullinger1980}&~~~~~~~~~~~~-&~1 072 083 327 (121)\\
\hline
Pickering~\cite{Pickering1998}&~1 069 338 652 (60)&~1 072 083 252 (60)\\
\hline
Aldenius~\cite{Aldenius2006}&~1 069 338 802 (60)&~1 072 083 342 (60)\\
\hline
\end{tabular}
\label{TableComposite}
\end{table}

The only isotopically resolved reference data stood alone for nearly three decades, when Drullinger et al.~\cite{Drullinger1980} measured the D$_{2}$ line in a Penning trap on a cloud of Mg$^+$ ions. The transition frequency of the $^{24}$Mg$^+$ component was measured against an iodine reference line with an uncertainty of 120~MHz and the $^{26-24}$Mg$^+$ isotope shift was measured to be 3050~(100)~MHz in agreement with our results (see table~\ref{TableResults},\ref{TableShift}). So far, there was no isotopically resolved reference data for the D$_1$ line.

The most accurate literature values for the D$_{1}$ and D$_{2}$ transition frequencies originate from isotopically unresolved Fourier transform (FT) spectroscopy, here an uncertainty of 60~MHz was reported for composite lines of natural isotope abundance ~\cite{Pickering1998,Pickering2000,Aldenius2006}. In order to compare isotopically resolved results to unresolved measurements we determine the center-of-gravity for a composition of natural isotope abundance ($^{24}$Mg:~78.99\%, $^{25}$Mg:~10\%, $^{26}$Mg:~11.01\%). According to Land\'{e}'s interval rule $\delta\nu^{25,24}$ represents the isotope shift between the center-of-gravity of the $^{25}$Mg$^+$ hyperfine structure multiplet and the $^{24}$Mg$^+$ component~\cite{Sobelman}, thus the center-of-gravity of the composite line is shifted by $0.1\times\delta\nu^{25,24}+0.1101\times\delta\nu^{26,24}$ with respect to the $^{24}$Mg$^+$ transition frequency. The low natural abundance of $^{25}$Mg reduces the uncertainty of the center-of-gravity frequency to about~2~MHz. 

Pickering~et~al.~\cite{Pickering1998,Pickering2000} measured the D$_1$ and D$_2$ transitions in the spectrum of a hollow-cathode discharge lamp. Isotope shifts and hyperfine constants reported in~\cite{Drullinger1980} were used for their line shape model fitted to the isotopically unresolved lines. The isotope shift measurement of the D$_2$ line was taken as an approximation for the D$_1$ line. While the transition frequencies deduced for the $^{24}$Mg$^+$ components are in reasonable agreement with our measurements (see table~\ref{TableResults}), the transition frequencies given for the composite lines significantly deviate (3~$\sigma$) from the results derived from our isotopically resolved measurement (see table~\ref{TableComposite}). We attribute this to discrepancies in the definitions of the composite line center. In~\cite{Pickering1998} the spacing between the $^{24}$Mg components and the center of the composite lines were reported to be 378~MHz~(D$_1$) and 360~MHz~(D$_2$), analyzing the center-of-gravity of isotopically resolved data we find 501~MHz~(D$_1$) and 502~MHz~(D$_2$, Drullinger~\cite{Drullinger1980}:~495~MHz). 

Transition frequencies reported in~\cite{Aldenius2006} were acquired by fitting a single Voigt profile to the composite line-shape, the composite line transition frequencies are in reasonable agreement with our center-of-gravity results~(see table~\ref{TableComposite}).

\subsection{Comparison with Isotope Shift Theory \label{section2b}}

\begin{table}[tb]
\caption{Comparison of measured and predicted $^{26-24}$Mg$^+$ isotope shifts. The field shift contributions are included (each +18~(20)~MHz limited by experimental nuclear charge radii~\cite{angeli2004}) and set the uncertainty level for the comparison of our measurement with theoretical investigations. Uncertainties for the SMS contribution are only given in~\cite{berengut2003}.}
\centering
\begin{tabular}{|l|l|l|}
\hline
Reference &~~~~D$_1$ [MHz] &~~~~D$_2$ [MHz] \\
\hline
\hline
Exp. this work~~~~~~~&~~3084.905 (93)~~&~~3087.560 (87)~~\\
\hline
Exp. Drullinger~\cite{Drullinger1980}~&~~~~~~~~~-  &~~3050 (100) \\
\hline
Theory Safronova~\cite{safronova2001}~~&~~3059  &~~3060   \\
\hline
Theory Tupitsyn~\cite{tupitsyn2003}~~&~~3200  &~~~~~~~~~-  \\
\hline
Theory Berengut~\cite{berengut2003}~~&~~3094 (43)  &~~3099 (28)  \\
\hline
Theory Korol~\cite{Korol2007}~~&~~3107  &~~3109  \\
\hline
\end{tabular}
\label{TableShift}
\end{table}

The accuracy of theoretical isotope shifts used to exceed experimental data~\cite{berengut2003,Drullinger1980}, our measurement reverses this situation. Theoretical isotope shifts from recent investigations~\cite{berengut2003,safronova2001,tupitsyn2003,Korol2007} are shown in table~\ref{TableShift}. Different results arise mainly from the determination of the SMS, since differences in the NMS and differences in the field shift due to uncommon field shift constants $F_{\rm{FS}}$ are marginal, compared to the uncertainty introduced by experimental rms nuclear charge radii~\cite{angeli2004}. Only the authors of~\cite{berengut2003} give error bars for their evaluation of the SMS constant {\it k}$_{\rm{SMS}}$, the uncertainty levels introduced by {\it k}$_{\rm{SMS}}$ and the nuclear charge radii are both on the order of 20~MHz. Therefore the considerably more precise value of our measurement can neither set an accurate calibration point for SMS investigations nor give rigorous insight on isotopic differences in the nuclear charge distribution. Yet the results of~\cite{berengut2003} and our measurement are in good agreement. Combining our results with the values given for {\it k}$_{\rm{SMS}}$ and $F_{\rm{FS}}$ from~\cite{berengut2003} allows to deduce the difference in the rms nuclear charge radii. Here we find $\delta\left\langle r^2\right\rangle^{24-26}$=~0.06~(15)~fm$^2$ from the D$_2$ line and $\delta\left\langle r^2\right\rangle^{24-26}$=~0.07~(30)~fm$^2$ from the D$_1$ result, consistent with the tabulated value of 0.14~(16)~fm$^2$~\cite{angeli2004}. 

The accuracy of our measurement allows to resolve the $^{26-24}$Mg$^+$ isotope shift of the forbidden 3p$_{3/2}$-3p$_{1/2}$ transition. Since the fine structure splitting is of purely relativistic origin our measurement may be an interesting test for relativistic effects in isotope shift investigations. One result of~\cite{Korol2007} was, that the final relativistic mass shifts for levels of a multiplet appear to be closer to each other than the mass shifts obtained with non relativistic operators, giving rise to a difference in the determined mass shifts of 2.2~MHz and -14.4~MHz for the relativistic and non relativistic treatment respectively (see table~V in~\cite{Korol2007}). Our measured value of 2.66~(13)~MHz for the total isotope shift supports this result.  
\\

\section{Summary and Outlook}
\label{section6}

We have accurately measured the $^{24}$Mg and $^{26}$Mg components of the 3s-3p fine structure doublet in Mg$^+$ using a new spectroscopy method for trapped ions in the weak binding limit~\cite{Herrmann2009}. Absolute frequency calibration (relative to the definition of the SI second) allows us to directly deduce the corresponding isotope shifts and fine structure splittings. The measurement accuracy exceeds previous results by more than two orders of magnitude and allows to resolve the isotope shift of the relativistic fine structure splitting for the first time. We deduce transition frequencies for the $^{25}$Mg hyperfine structure and the center-of-mass of an isotope mixture of natural abundance. A further measurement of the $^{25}$Mg component would improve the uncertainty of composite transition frequencies, isotope shifts and hyperfine constants for the 3p$_{1/2}$ and 3p$_{3/2}$ level~\cite{Safronova1998,Sur2005}. In turn it would allow for a more stringent consistency check of theoretical hyperfine structure constants, isotope shift theories and nuclear charge distributions.

Though our measurement may be regarded as exact reference for present and near future astrophysical spectra and theoretical isotope shifts, future applications may require even more accurate calibration. Narrowing down the linewidth of the spectroscopy laser e.g.~by stabilization to an ultra-stable reference cavity would essentially remove the limiting systematic effect of our measurement. The statistical uncertainty of the $^{26}$D$_2$ line was only 14~kHz, indicating that our spectroscopy scheme may allow for further improvement of at least one order of magnitude.     
 
This research was supported by the DFG cluster of excellence
``Munich Centre for Advanced Photonics''.


\begin{thebibliography}{10}


\bibitem{Bahcall1967}
J. N. Bahcall, W. L. W. Sargent, and M. Schmidt, ApJ {\bf149}, L11, (1967).
  
\bibitem{Wolfe}
A. M. Wolfe, R. L. Brown, and M. S. Roberts, Phys. Rev. Lett. {\bf37}, 179 (1976).

\bibitem{Savedoff1956}
M. P. Savedoff, Nature {\bf178}, 688 (1956).

\bibitem{Dzuba99}
V. A. Dzuba, V. V. Flambaum, and J. K. Webb, Phys. Rev. Lett. {\bf82}, 888 (1999).

\bibitem{Webb99}
J. K. Webb, V. V. Flambaum, C. W. Churchill, M. J. Drinkwater, and J. D. Barrow, Phys. Rev. Lett. {\bf82}, 884 (1999).

\bibitem{Murphy2001}
 M. T. Murphy, J. K. Webb, P V. V. Flambaum, V. A. Dzuba, C. W. Churchill, J. X. Prochaska, J. D. Barrow and A. M. Wolfe, Mon. Not. Roy. Astron. Soc. {\bf327}, 1208 (2001).

\bibitem{Murphy2003}
M. T. Murphy, J. K. Webb, V. V. Flambaum, Mon. Not. Roy. Astron. Soc. {\bf345}, 609 (2003).

\bibitem{Murphy2004}
M. T. Murphy, V. V. Flambaum, J. K. Webb, V. V. Dzuba, J. X. Prochaska and A. M. Wolfe, Lecture Notes Phys. {\bf648}, 131 (2004).

\bibitem{keckvlt}
For further information see www.keckobservatory.org and www.eso.org
  
\bibitem{Srianand2004}
R. Srianand, H. Chand, P. Petitjean and B. Aracil, Phys. Rev. Lett. {\bf92},
121302 (2004). Full description: Chand et al., e-print arXiv:astro-ph/0401094v1

\bibitem{Murphy2008}
M. T. Murphy, J. K. Webb, V. V. Flambaum, Mon. Not. Roy. Astron. Soc. {\bf384}, 1053 (2008).  

\bibitem{Molaro2008}
P. Molaro, Eur. Phys. J. Special Topics {\bf163}, 173 (2008).

\bibitem{Molaro2007}
P. Molaro, e-print arXiv:0712.4390v1

\bibitem{Murphy2007}
M. T. Murphy, Th. Udem, R. Holzwarth, A. Sizmann, L. Pasquini, C. Araujo-Hauck, H. Dekker, S. D'Odorico, M. Fischer, T. W. H\"{a}nsch, and A. Manescau, Mon. Not. Roy. Astron. Soc. {\bf380}, 839 (2007).

\bibitem{Steinmetz2008}
T.~Steinmetz, T. Wilken, C. Araujo-Hauck, R. Holzwarth, T. W. H\"{a}nsch, L. Pasquini, A. Manescau, S. D'Odorico, M. T. Murphy, T. Kentischer, W. Schmidt, and Th. Udem, Science {\bf321}, 1335 (2008).

\bibitem{Braje2008}
D. A. Braje, M. S. Kirchner, S. Osterman, T. Fortier, and S. A. Diddams, Eur. Phys. J. D {\bf48}, 57 (2008)

\bibitem{Li2008}
C. H. Li, A. J. Benedick, P. Fendel, A. G. Glenday, F. X. K\"{a}rtner, D. F. Phillips, D. Sasselov, A. Szentgyorgyi, and R. L. Walsworth, Nature {\bf452}, 610 (2008).

\bibitem{Berengut2006}
J.~C.~Berengut, V. A. Dzuba, V. V. Flambaum, M. G. Kozlov, M. V. Marchenko, M. T. Murphy, and J. K. Webb, e-print arXiv:physics/0408017 (2006).

\bibitem{comb}
S. A. Diddams, D. J. Jones, J. Ye, S. T. Cundiff, J. L. Hall, J. K. Ranka, R. S. Windeler, R. Holzwarth, T. Udem, and T. W. H\"{a}nsch, Phys. Rev. Lett. {\bf84}, 5102 (2000). 

\bibitem{Wolf2008}
A. L. Wolf, S. A. van den Berg, C. Gohle, E. J. Salumbides, W. Ubachs, and K. S. E. Eikema, Phys. Rev. A {\bf78}, 032511 (2008).

\bibitem{Wolf2009}
A. L. Wolf, S. A. van den Berg, W. Ubachs, and K. S. E. Eikema, Phys. Rev. Lett. {\bf102}, 223901 (2009).

\bibitem{Herrmann2009}
M. Herrmann, V. Batteiger, S. Kn\"{u}nz, G. Saathoff, Th. Udem, and T. W. H\"{a}nsch, Phys. Rev. Lett. {\bf102}, 013006 (2009).

\bibitem{Kozlov2004}
M.G.~Kozlov, V. A. Korol, J. C. Berengut, V. A. Dzuba, and V. V. Flambaum, Phys. Rev. A {\bf70}, 062108 (2004).

\bibitem{Ashenfelter2004}
T. Ashenfelter, G. J. Mathews, and K. A. Olive, Phys. Rev. Lett. {\bf92}, 041102 (2004).

\bibitem{Fenner2005}
Y. Fenner, M. T. Murphy, and B. K. Gibson, Mon. Not. Roy. Astron. Soc. {\bf358}, 468 (2005).

\bibitem{berengut2003}
J. C. Berengut, V. A. Dzuba, and V. V. Flambaum, Phys. Rev. A {\bf68}, 022502 (2003).

\bibitem{Drullinger1980}
R. E.~Drullinger, D.~J.~Wineland, and J. C.~Bergquist, Appl. Phys. {\bf22}, 365 (1980). 

\bibitem{Pickering1998}
J. C.~Pickering, A. P. Thorne, and J. K. Webb, Mon. Not. Roy. Astron. Soc. {\bf300}, 131 (1998).

\bibitem{Pickering2000}
J. C.~Pickering, A. P. Thorne, J. E. Murray, U. Litzen, S. Johansson, V. Zilio, and J. K. Webb, Mon. Not. Roy. Astron. Soc. {\bf319}, 163 (2000).

\bibitem{Aldenius2006}
M.~Aldenius, S. Johansson, and M. T. Murphy, Mon. Not. Roy. Astron. Soc. {\bf370}, 444 (2006).

\bibitem{Rosenband2008}
T. Rosenband, D. B. Hume, P. O. Schmidt, C. W. Chou, A. Brusch, L. Lorini, W. H. Oskay, R. E. Drullinger, T. M. Fortier, J. E. Stalnaker, S. A. Diddams, W. C. Swann, N. R. Newbury, W. M. Itano, D. J. Wineland, and J. C. Bergquist, Science {\bf319}, 1808 (2008).

\bibitem{Bergquist1987}
J. C. Bergquist, W. M. Itano, and D. J. Wineland, Phys. Rev. A {\bf 36}, 428 (1987). 

\bibitem{Wineland1979}
D.~J. Wineland and W.~M. Itano, Phys. Rev. A {\bf 20}, 1521 (1979).

\bibitem{Nagourney1983}
W. Nagourney, G. Janik, and H. Dehmelt, Proc. Natl. Acad. Sci. USA {\bf80}, 643 (1983).

\bibitem{Wineland1981}
D. J. Wineland, and W. M. Itano, Phys. Lett. {\bf82A}, 75 (1981).

\bibitem{Nakamura2006}
T.~Nakamura, et al. Phys. Rev. A {\bf 74}, 052503 (2006).

\bibitem{Kjaergaard2000}
N. Kjaergaard, L. Hornekaer, A. M. Thommesen, Z. Videsen, and M. Drewsen, Appl. Phys. B {\bf71}, 207 (2000).
 
\bibitem{Gulde2001}
S. Gulde, D. Rotter, P. Barton, F. Schmidt-Kaler, R. Blatt, and W. Hogervorst, Appl. Phys. B {\bf73}, 861 (2001).

\bibitem{Lucas2004}
D. M. Lucas, A. Ramos, J. P. Home, M. J. McDonnell, S. Nakayama, J. P. Stacey, S. C. Webster, D. N. Stacey, and A. M. Steane, Phys. Rev. A {\bf69}, 012711 (2004).

\bibitem{Steele2007}
A. V. Steele, L. R. Churchill, P. F. Griffin, and M. S. Chapman, Phys. Rev. A {\bf75}, 053404 (2007).

\bibitem{Tanaka2007}
U. Tanaka, I. Morita, and S. Urabe,  Appl. Phys. B {\bf89}, 195 (2007).

\bibitem{Salumbides2006}
E. J. Salumbides, S. Hannemann, K. S. E. Eikema and W. Ubachs, Mon. Not. Roy. Astron. Soc. {\bf373}, L41 (2006).

\bibitem{Friedenauer2006}
A. Friedenauer, F. Markert, H. Schmitz, L. Petersen, S. Kahra, M. Herrmann, Th. Udem, T. W. H\"{a}nsch and T. Sch\"{a}tz, Appl. Phys. B {\bf 84},  371  (2006).

\bibitem{DPAOM}
E. A. Donley, T. P. Heavner, F. Levi, M. O. Tataw, and S. R. Jefferts, Rev. Sci. Instrum. {\bf76}, 063112 (2005).

\bibitem{Drake}
G. W. F. Drake, {\itshape Handbook of Atomic, Molecular, and Optical Physics}, Springer (2006).

\bibitem{safronova2001}
M. S. Safronova, and W. R. Johnson, Phys. Rev. A {\bf64}, 052501 (2001).

\bibitem{tupitsyn2003}
I. I. Tupitsyn, V. M. Shabaev, J. R. Crespo Lopez-Urrutia, I. Draganic, R.
Soria Orts, and J. Ullrich, Phys. Rev. A {\bf68}, 022511 (2003).

\bibitem{Korol2007}
V. A. Korol and M. G. Kozlov, Phys. Rev. A {\bf76}, 022103 (2007).

\bibitem{angeli2004}
I. Angeli, Atomic Data and Nuclear Data Tables {\bf87}, 185 (2004). 

\bibitem{Itano1981}
W. M.~Itano and D.~J. Wineland, Phys. Rev. A {\bf 24}, 1364 (1981).

\bibitem{Sur2005}
C. Sur, B. K. Sahoo, R. K. Chaudhuri, B. P. Das, and D. Mukherjee, Eur. Phys. J. D {\bf 32}, 25 (2005)

\bibitem{Safronova1998}
M. S. Safronova, A. Derevianko, and W. R. Johnson, Phys. Rev. A {\bf 58}, 1016 (1998)

\bibitem{Sobelman}
I. I.~Sobelman, {\itshape Atomic Spectra and Radiative Transitions}, Springer (1996). 

\end{thebibliography}
\end{document}